\documentclass[
    ,final            
  ]
  {aipproc}

\layoutstyle{6x9}

\begin{document}

\title{Producing and Detecting Correlated Atoms}


\author{C. I. Westbrook}{
  address={Laboratoire Charles Fabry de l'Institut d'Optique, F-91403 Orsay Cedex, France}}
\author{M. Schellekens}{
  address={Laboratoire Charles Fabry de l'Institut d'Optique, F-91403 Orsay Cedex, France}}
  \author{A. Perrin}{
  address={Laboratoire Charles Fabry de l'Institut d'Optique, F-91403 Orsay Cedex, France}}
\author{V. Krachmalnicoff}{
  address={Laboratoire Charles Fabry de l'Institut d'Optique, F-91403 Orsay Cedex, France}}
\author{J. Viana Gomes}{
  address={Laboratoire Charles Fabry de l'Institut d'Optique, F-91403 Orsay Cedex, France}
    ,altaddress={Departamento de Fisica, Universidade do Minho, Campus
de Gualtar, 4710-057 Braga, Portugal}}
  \author{J.-B. Trebbia}{
  address={Laboratoire Charles Fabry de l'Institut d'Optique, F-91403 Orsay Cedex, France}}
\author{J. Est\`eve}{
  address={Laboratoire Charles Fabry de l'Institut d'Optique, F-91403 Orsay Cedex, France}}
\author{H. Chang}{
  address={Laboratoire Charles Fabry de l'Institut d'Optique, F-91403 Orsay Cedex, France}}
  \author{I. Bouchoule}{
   address={Laboratoire Charles Fabry de l'Institut d'Optique, F-91403 Orsay Cedex, France}}
   \author{D. Boiron}{
   address={Laboratoire Charles Fabry de l'Institut d'Optique, F-91403 Orsay Cedex, France}}
   \author{A. Aspect}{
  address={Laboratoire Charles Fabry de l'Institut d'Optique, F-91403 Orsay Cedex, France}}

\author{T. Jeltes}{
  address={Laser Center Vrije Universiteit, 1081 HV Amsterdam, the Netherlands}}
  \author{J. McNamara}{
  address={Laser Center Vrije Universiteit, 1081 HV Amsterdam, the Netherlands}}
    \author{W. Hogervorst}{
  address={Laser Center Vrije Universiteit, 1081 HV Amsterdam, the Netherlands}}
\author{W. Vassen}{
  address={Laser Center Vrije Universiteit, 1081 HV Amsterdam, the Netherlands}}


\begin{abstract}
We discuss experiments to produce and detect atom correlations in a degenerate or nearly degenerate gas of neutral atoms. First we treat  the atomic analog of the celebrated Hanbury Brown Twiss experiment, in which atom correlations result simply from interference effects without any atom interactions. We have performed this experiment for both bosons and fermions. Next we show how atom interactions produce correlated atoms using  the atomic analog of spontaneous four-wave mixing. Finally, we briefly mention experiments on a one dimensional gas on an atom chip in which correlation effects due to both interference and interactions have been observed.
\end{abstract}

\maketitle


\section{Introduction}

Recent years have seen a blossoming in the use of experimental techniques sensitive to atom correlation in the study of ultra-cold atomic gases. After a pioneering experiment in 1996 \cite{Yasuda:96}, in which the atomic Hanbury Brown Twiss (HBT) experiment was first observed, there were many analyses and proposed extensions to other situations \cite{naraschewski:99, cahill:99, Grondalski:99,Altman:04}. In the past two years several experimental realizations have been reported \cite{Foelling:05,Oettl:05,Greiner:05,Schellekens:05,Esteve:06}. At the same time, the theoretical community has shown much interest in correlated pairs of atoms, either from collisions or from the breakup of molecules\cite{ Duan:00, Moore:00, Zin:06,Norrie:06,Kheruntsyan:05,Deuar:06}. Here again, the year 2005 saw the report of an experimental realization. In this paper we will discuss some new experiments concerning atom correlations. We refer the reader to the contribution of I. Bloch in this volume for additional ones. All this activity promises to provide much new information about the behavior of cold quantum gases, but we also emphasize  that in the field of nuclear and particle physics, Hanbury Brown Twiss correlations are a well established experimental technique and we recomend Ref.~\cite{Boal:90} for a review.

\section{The atomic Hanbury Brown Twiss experiment}

\subsection{Intuitive picture}

At this conference, the contributions of R. Glauber and that of E. Demler discuss the theoretical interpretation of the correlation experiments. Here we will give a less general but intuitive point of view due to Fano \cite{Fano:61}. The HBT effect necessarily involves the detection of two particles at different space-time points. Thus one is led to consider, two source points, A and B, which emit particles detected at two detection points,
$C$ and $D$. One must consider the quantum mechanical amplitude
for the process ($A\rightarrow C$ and $B \rightarrow  D$) as well
as that for ($A \rightarrow  D$ and $B \rightarrow  C$). If the
two processes are indistinguishable, the amplitudes interfere. For
bosons, the interference is constructive resulting in a joint
detection probability which is enhanced compared to that of two
statistically independent detection events, while for fermions the
joint probability is lowered. For a detector separation larger than the aperture which 
would permit the resolution of the structure of the source, the average over different points in the
source washes out the interference and one recovers
the situation for uncorrelated particles.

In the case of a chaotic source of light the correlations are also easily understood in terms of speckle, without reference to the concept of photons. But, as in optics with light, we are capable of producing sources for which a classical analysis is not adequate. The experiment discussed below using fermions is an example. The idea of speckle can still be useful however, because it reminds us that these experiments can be analyzed from the point of view of noise or fluctuation phenomena. 

\subsection{Experiment with bosons}

The detection of HBT correlations has presented a significant experimental challenge, because the correlation is only maximal if the two detectors occupy the same $\hbar^3$ volume in phase space. In other words, the correlation length, ${\hbar t}\over{m s}$ where $s$ is the size of the source and $t$ is the time of flight to the detector\cite{Gomes:06}, is generally quite small. In addition, the signal to noise ratio in the experiment is proportional to the phase space density of the source.  Thus, the detailed study of the HBT effect was greatly facilitated by the advent of degenerate quantum gases. Figure~\ref{detector} shows our detector. A micro-channel plate (MCP) amplifies the electron ejected upon impact by a metastable helium atom (20 eV internal energy). The charge pulse is recorded with ns time resolution and in the horizontal plane a delay line anode and timing electronics provides position resolution\footnote{Available from Roentdek \url{http://www.roendtek.com}. Our time to digital converter is manufactured by ISITech \url{http://www.opticsvalley.org/data/isitech.pdf}}. Although the detector actually measures arrival times, we conventionally convert the arrival time into a vertical position by multiplying by the velocity at arrival at the detector. This velocity has a spread of less than 1\% over the atomic sample.

In the spring of 2005, we successfully used this detector to observe the HBT effect using an evaporatively cooled sample of $^4$He\cite{Schellekens:05,Westbrook:05}. We refer the reader to those papers for more information. The shape, width and height of the correlation signal can be quantitavely understood by a simple ideal gas treatment of the atoms. 

\subsection{Experiment with fermions}

\begin{figure}
  \includegraphics[height=.35\textheight]{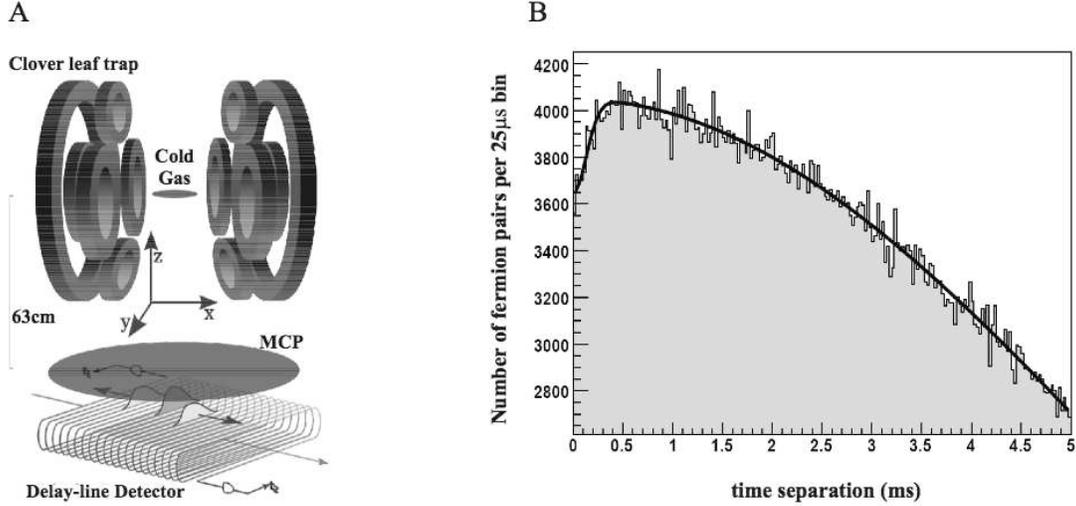}
  \caption{(A). Drawing of the cloverleaf trap and the position sensitive detector. The detector consists of two 8 cm diameter micro channel plates. The charge is collected by two delay-line anodes to give both the arrival time (or equivalently the vertical position) and the position of the particles in the x-y plane. The vertical resolution is determined by the time resolution (1 ns)  while the horizontal resolution is about 500~$\mu$m. (B). Pair distribution histogram for $^3$He falling on the detector. A time separation of 1 ms corresponds to about 3.5 mm. The broad overall shape of the distribution (HWHM about 6 ms) is due to the approximately Gaussian temporal shape of the cloud. The dip for times below 1 ms is the antibunching or Fermibury effect.} 
  \label{detector}
\end{figure}

Only weeks before the ICAP2006 conference, the metastable helium groups in Orsay and in Amsterdam began a collaboration to  observe the analagous effect with the fermionic isotope $^3$He. Instead of a bump, we expect a dip. The Amsterdam group had already demonstrated the production of a degenerate gas of $^3$He \cite{McNamara:06}, and the setup was sufficiently similar to the one in Orsay that it was possible to install  the Orsay detector in the Amsterdam apparatus with some minor modifications to the vacuum system. It took only two weeks of (albeit intense) work to see an unambiguous anti-bunching signal. One of the first of these is shown in Fig.~\ref{detector}. Unlike the published boson data, the data shown are not normalized. Thus one sees a broad structure in the peak corresponding to the Gaussian shape of the cloud as it arrives at the detector (or more precisely its auto-convolution). The anti-bunching signal is the small dip for pair separations below 1 ms. At this writing, quantitative analysis of these data is still in progress, but roughly speaking, the antibunching signal corresponds closely to our expectations. Under identical conditions, the width and amplitude of the signal for bosons and fermions are of similar magnitude. 

\section{Observation of correlated atom pairs}

Interactions between atoms also produce correlated atoms. The correlations are particularly simple for elastic collisions, and obviously the study of collision products has occupied much of atomic, nuclear and particle physics for the last century. The collisions we study here however are a little different because the source is a Bose-Einstein condensate (BEC), and therefore approximates a "single mode" source. Thus, analogies with four-wave mixing are very apt. Indeed stimulated four wave mixing using BEC's has already been observed\cite{Deng:99,Vogels:02}. Collisions between BEC's have also been studied by two other groups\cite{Buggle:04,Thomas:04}. In these experiments collision velocities were sufficiently high that not only s-waves but also d-waves were involved. The experiments we describe here are limited to the s-wave regime. 

The point of departure for the experiment is the Orsay apparatus used in the HBT experiment described above. To this apparatus we have added two more laser beams to drive stimulated Raman transitions between the trapped ($m=1$) and field insensitive ($m=0$) states of the $2^3S_1$ energy level.  The quantization axis is defined by the magnetic field which is along the x-axis. The two lasers are detuned by 400 MHz from the  $2^3S_1$ -  $2^3P_0$ transition and have a relative detuning of about 600 kHz in order to be in Raman resonance for atoms in the bias field of the magnetic trap. The laser beams propagate at small angles to the $z$ and $x$ axes of Fig. 1 and thus a momentum transfer of $\hbar k(\mathbf e_x+{\mathbf e_z})$, where $\mathbf e_x$ is a unit vector along the x-axis, accompanies the Raman transition. The beam along the $x$-axis is retro-reflected resulting in a second possible Raman transition with a momentum transfer of $\hbar k(-\mathbf e_x+{\mathbf e_z})$.

\begin{figure}
  \includegraphics[height=.5\textheight]{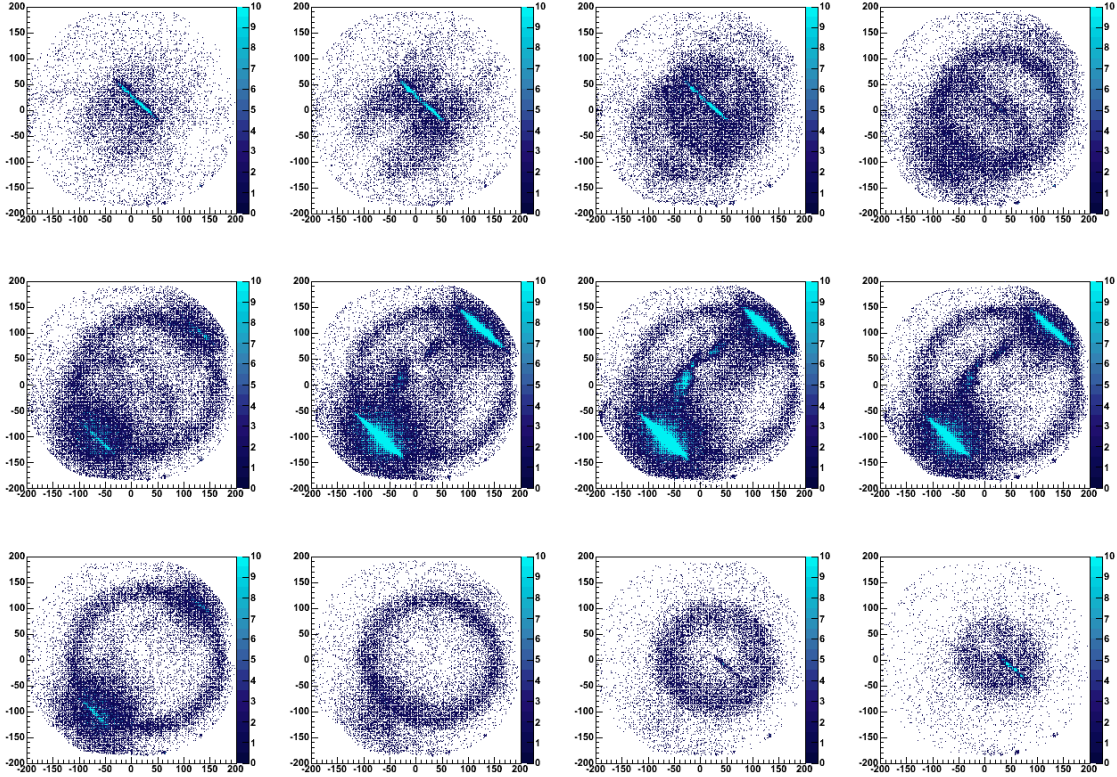}
  \caption{Images of the collision of two condensates. Each frame represents successive slices of the atomic cloud as it passes the plane of the detector. The two colliding condensates and the s-wave collision sphere are clearly visible.} 
  \label{circle}
\end{figure}

Most of the atoms are transfered to the $m=0$ state with one of the two possible momentum components. These atoms collide with a relative velocity of twice the recoil velocity and, in the absence of gravity, would be scattered into a spherical shell whose radius is ${\hbar k \over m}t$ where $t$ is the time after the collision. In the presence of gravity, the kinematics are the same except that the spherical shell accelerates downward. Atoms which are not transferred to the $m=0$ state remain in the trap. Figure~\ref{circle} shows some data taken with the delay line detector under these conditions. The figure shows successive slices of the three-dimensional reconstruction of the atoms' positions. The sphere into which the atoms are scattered, as well as the two colliding clouds are clearly visible. Less than 10\% of the atoms are scattered from the colliding clouds. 

\begin{figure}
  \includegraphics[height=.45\textheight]{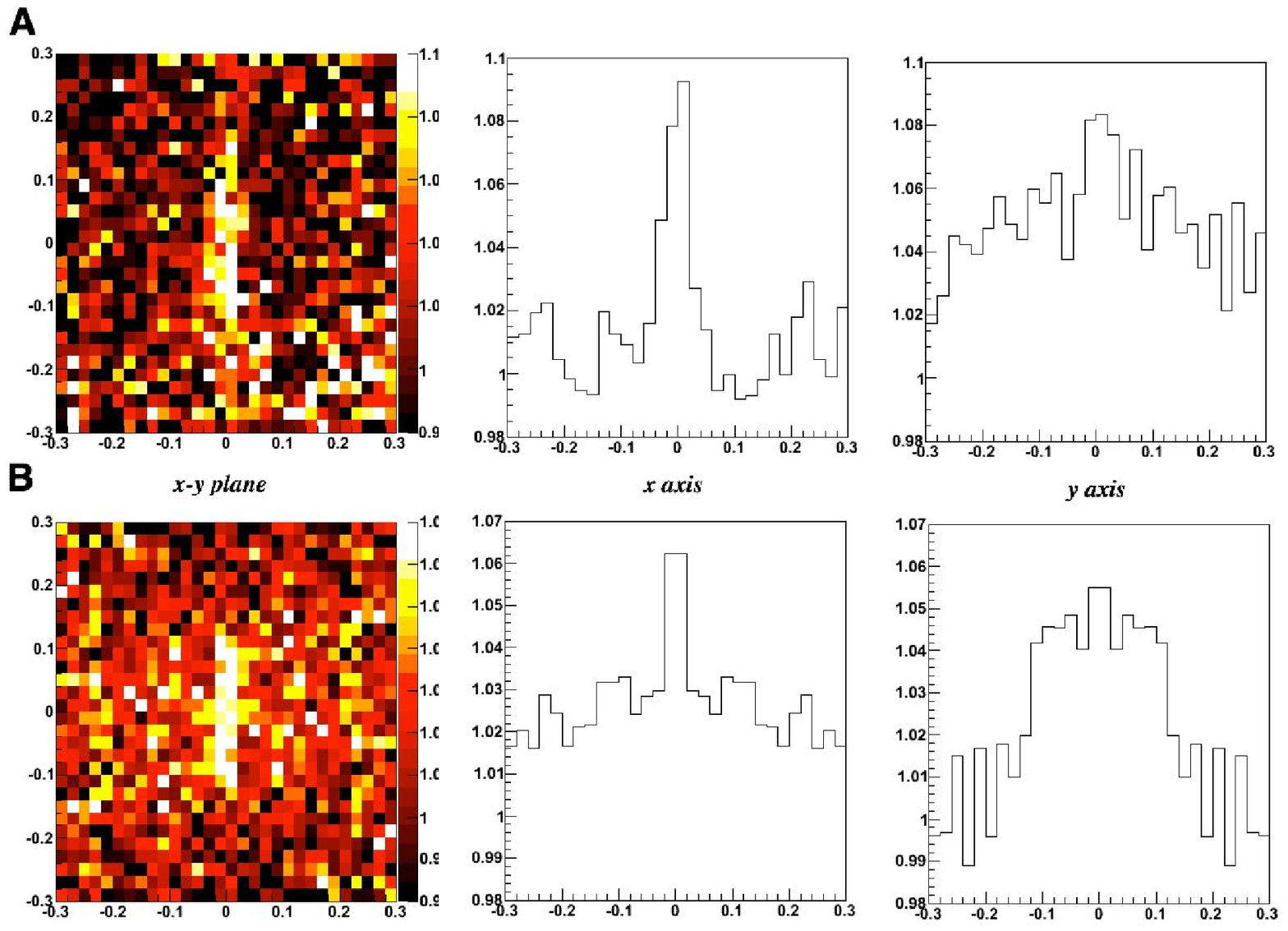}
  \caption{Atom correlation signals: (A), for opposite momenta, (B), for colinear momenta. The left image shows the signal in the x-y plane. The next two in each row show the signals along one axis and integrated along the other.  The second line (B) is simply a manifestation of the Hanbury Brown Twiss effect. In both cases the correlation function is anisotropic because of the initial anisotropy of the source. The normalization is such that unity corresponds to no correlation.} 
  \label{pi-corr}
\end{figure}

The correlations on the sphere are shown in Fig.~\ref{pi-corr}.  We see a correlation signal both for pairs of opposite momenta in the center of mass frame, $(p,-p)$ as well as for pairs with the same momentum $(p,p)$. The latter effect is simply another manifestation of the HBT effect\cite{Zin:06,Deuar:06} and may prove useful because it allows us to characterize the size of the source. The size of signal for opposite momenta can in principle be many times the background level. Its small size here appears to be due to a rather poor quantum efficiency of the detector (on the order of 5\% averaged over the detector). The width of the anti-colinear correlation peak is clearly larger than the colinear peak. We believe that its width is partly due to the mean field energy acquired by the atoms during their expansion\cite{Zin:06,Norrie:05,Kheruntsyan:05,Deuar:06}. A quantitative study of the width and the detector quantum efficiency is in progress.

\section{Correlations on an atom chip}

As has been shown in other recent experiments, single atom counting is not necessary to observe atom correlation effects. Absorptive imaging, when performed in sufficiently low noise conditions can also be sensitive to atom number or atom density fluctuations\cite{Altman:04,Foelling:05, Greiner:05}. Inspired by these experiments, we have also examined number fluctuations in a nearly one dimensional gas on an atom chip. A difficulty in absorption imaging is the fact that the necessary integration over one direction can average out the desired signal.  The one dimensional geometry is particularly favorable in this respect because this integration can be avoided. Atom chips are also advantageous for this sort of study because they permit the use of a compact, and mechanically stable apparatus. In our experiment we had little difficulty in taking images at the photon shot-noise limit. Multiple averages (several hundreds) of these images permitted an accurate subtraction of the photon shot-noise to reveal the atom number fluctuations\cite{Esteve:06}.  

In spite of the fact that the resolution of our imaging system was significantly larger than the correlation length of the sample, these measurements had two new features. First, the fluctuations were observed without releasing the atoms from the trap. They were thus sensitive to correlations in position rather than momentum. Second, and more importantly, we were able to identify a regime of high density in which interactions between the atoms suppressed the fluctuations which would be expected for a non-interacting gas. A careful analysis of the density profile of such a gas also reveals that the profile cannot be explained by a Hartree-Fock calculation that neglects correlations between the particles \cite{Trebbia:06}. Improved calculations, including correlation effects promise to give a better account of the observations \cite{Blakie:05, Proukakis:06}.\\

We hope that the experiments discussed above  have given a taste of the rich possibilities in correlation measurements. The main conclusion with which we would like to leave the reader is that, in the field of degenerate quantum gases, treating correlations between particles remains an important challenge. But our  increasingly sophisticated experimental techniques are beginning provide a window on these phenomena, and we can expect great progress in the near future.


\begin{theacknowledgments}
The Laboratoire Charles Fabry is a member of the CNRS Federation LUMAT (FR2764), and of the Institut Francilien pour la Recherche en Atomes Froids (IFRAF). We acknowledge the technical help provided by the DPTI Technology Platform of the CNRS and the Universit\'e de Paris-sud. We acknowledge the support of the EU under grants MRTN-CT-2003-505032 (Atom Chips) and IP-CT-015714 (SCALA) and of the ANR under contract 05-NANO-008-01.The Amsterdam/Orsay collaboration is supported by the European Community Integrated Infrastructure Initiative action (RII3-CT-2003-506350) and by the Netherlands Foundation for Fundamental Research of Matter (FOM).
 \end{theacknowledgments}



\bibliographystyle{aipproc}   





\end{document}